\documentclass[twocolumn,epsf,prb,amsmath,amssymb,showpacs,floatfix]{revtex4}
\usepackage{graphicx}
\usepackage{amsmath}
\usepackage{amssymb}
\usepackage{color}
\usepackage[T1]{fontenc}
\newcommand{\nup}[1]{\hat{n}_{#1 \sigma}}

\newcommand{\ciup}[1]{\hat{c}_{#1 \sigma}}
\newcommand{\cdgiup}[1]{{\hat{c}^\dag}_{#1 \sigma}}
\newcommand{\cdgidw}[1]{{\hat{c}^\dag}_{#1 \bar{\sigma}}}

\newcommand{\ndw}[1]{\hat{n}_{#1 \bar{\sigma}}}
\newcommand{\nupzr}[1]{{\langle \hat{n}_{#1 \sigma} \rangle}_0}
\newcommand{\ndwzr}[1]{{\langle \hat{n}_{#1 \bar{\sigma}} \rangle}_0}
\newcommand{\qup}[1]{q_{#1 \sigma}}

\begin{document}
\title{Field effect on surface states in a doped Mott-Insulator thin film}
\author{D. Nasr Esfahani}
\email{Davoud.NasrEsfahani@ua.ac.be}
\author{L. Covaci}
\email{lucian@covaci.org} 
\author{F. M. Peeters}
\email{Francois.Peeters@ua.ac.be}
\affiliation{Departement Fysica, Universiteit Antwerpen,
Groenenborgerlaan 171, B-2020 Antwerpen, Belgium}
\begin{abstract}
Surface effects of a doped thin film made of a strongly correlated material are investigated both in the absence and presence of a perpendicular electric field. We use an inhomogeneous Gutzwiller approximation for a single band 
Hubbard model in order to describe correlation effects. For low doping, the bulk value of the quasiparticle weight is recovered exponentially deep into the slab, but with increasing doping, additional Friedel oscillations appear
near the surface. We show that the inverse correlation length has a power-law dependence on the doping level. In the presence of an electrical field, considerable changes in the quasiparticle weight can be realized throughout the
system. We observe a large difference (as large as five orders of magnitude) in the quasiparticle weight near the opposite sides of the slab. This effect can be significant in switching devices that use the surface states for 
transport.
\end{abstract}
\maketitle
\section{Introduction}
The metal insulator transition (MIT) based on carrier doping of a Mott insulator has been investigated experimentally and theoretically\cite{georges,kotliar}. Recently, the formation of a superconducting phase was observed at the
interface of a Mott and band insulator and the possible tuning of these transitions by an external electric field was reported\cite{caviglial}. Moreover a three terminal setup was implemented by Son et al. who induced hole doping
in a thin Mott insulator film from a doped band insulator through the application of a voltage difference between the drain and the gate terminals\cite{son}.\\
For the above class of phenomena inhomogeneities and proximity effects play an essential role. In order to deal with such systems one needs a theoretical model that is able to include correlation effects in heterostructures
while not being too computationally expensive such that one has the possibility to consider large enough system sizes. This is crucial especially for the investigation of systems where surfaces and finite size effects are significant such as thin films made of strongly correlated materials.
The interface between a band insulator and a strongly correlated system has been studied theoretically with a two site dynamical mean field theory(DMFT)\cite{millis} and the slave boson mean field theory(SBMFT)\cite{anderas}. 
Such studies predict the formation of a two dimensional electron gas at the interface which arises due to charge reconstruction.  Surface correlation effects were studied theoretically in half filled heterostructures modeled 
by a single band Hubbard model\cite{potthof}. Also the penetration of metallic behavior into a Mott insulator was studied both within the Gutzwiller approximation and DMFT for the half filled case\cite{costi,michele}. Surface
correlation effects of a doped semi-infinite Hubbard model were investigated within an embedded DMFT for both single band and multi-band systems\cite{nourafkan,nourafkan2}. Within this method, due to numerical limitations, 
only few surface layers (up to 6) can be used in order to address site dependent correlation effects. When the correlation length is large, this method is not reliable any more.\\
In order to describe position dependent electronic correlation effects in a slab geometry we employ the Gutzwiller approximation(GA). While GA works only for the metallic phase, it gives reliable information about the quasiparticle
(QP) weight of electrons at different spatial locations. For heterostructures, GA was found to be in good qualitative agreement with the more refined DMFT method for the half filled case\cite{michele}. While GA and SBMFT are 
equivalent for zero temperature\cite{bunemannslave}, in two site DMFT, like GA, the bulk QP weight is governed by a simple power law and there is only a correction to $U_c$ when compared with the linearized DMFT\cite{zhang,potthof1}.
Generally, GA over-estimates the QP weight at all dopings but it is considered to be accurate enough to describe low energy excitations and is routinely used for interpolations in combination with DMFT methods\cite{kotliar}.\\
The aim of this paper is to investigate the spatial dependence of the charge density and the QP weight of a doped correlated slab and to understand the correlation effects in the presence of an external electric field. We predict 
significant changes in the QP weight throughout the system. This study is motivated by potential applications in nanoscale switching devices with spatial controllable conductivity through the application of an external electric field.\\
The outline of the paper is as follows: after a brief derivation of the saddle point equations for a slab geometry(section II) the results for a doped correlated slab are presented in section III(A). Next the effect of 
an electric field is discussed in section III(B) and finally we present our conclusions in section IV.
\section{Model and Method}
The simplest Hamiltonian that is able to capture the essential physics of strongly correlated systems is the single band Hubbard model\cite{hubbard},
\begin{equation}
 \hat{H}_U = -\sum_{\langle ij \rangle \sigma} t_{ij} c^\dagger_{i\sigma}c_{j\sigma}  + \sum_i U \nup i\ndw i ,
\end{equation}
where $t_{ij}$ are the hopping amplitudes, $\langle ij \rangle$ is summation over nearest neigthbour sites and $U$ is the Hubbard energy describing the Coulomb interaction between two electrons with opposite spin located on the same site. In the presence of an external electric field the model becomes \cite{graf}:
\begin{equation}
 \hat{H} = \hat{H}_U + \sum_{i\sigma}  v_i\hat{n}_{i\sigma},
\label{eq:H}
\end{equation}
where $v_i$ is the position dependent potential.
In spite of the simple form of the Hubbard model, exact solutions exist only for $d=1$ and $d=\infty$\cite{lieb,MV1,georges} and therefore we are forced to work with approximations. If one is only concerned about ground state 
properties or low energy excitations\cite{bunemann}, one of the choices is the Gutzwiller approximation(GA) which is the infinite dimension limit of the Gutzwiller wave function (GWF)\cite{gutz,MV1,geb}. GWF is a many body wave 
function with an additional degree of freedom used to reduce the weight of higher energy configurations. In the single band Hubbard model these configurations are on-site double occupancies obtained when two electrons with 
opposite spin reside on the same site. The GWF is written as:
\begin{equation}
|GWF\rangle = \prod_{i} \hat{P}_i |\phi_0\rangle,
\end{equation}
where $i$ is the lattice site index and the projector operators are defined as $\hat{P}_i = g_{e,i} \hat{e}_i + g_{\sigma,i} \hat{s}_{\sigma,i} + g_{\bar{\sigma},i} \hat{s}_{\bar{\sigma},i} + g_{d,i} 
\hat{d}_i$. The operators $\hat{e}=(1-\nup i )(1-\ndw i), \hat{s}_\sigma=\nup i (1-\ndw i )$ and $\hat{d}=\nup i \ndw i$
are local projectors of zero, singly and doubly occupied states, $|\varphi_0\rangle$ is a noninteracting Fermi sea and consist of both spin up 
 and spin down states and the $g$ coefficients are variational parameters. The following local constraints have to be satisfied in order to remove the local contributions in the diagramatic expansion of various expectation 
 values\cite{geb,bunemann},
\begin{eqnarray}
 &&\langle\hat{P}^\dag_i\hat{P}_i \rangle_0 = 1,\\
 &&\langle\hat{P}^\dag_i \hat{P}_i c^\dag_{i\sigma} c_{i\sigma} \rangle_0 = \langle c^{\dag}_{i\sigma} c_{i\sigma} \rangle_0.
\end{eqnarray}
Where $\langle \dots \rangle_0$ represents the expectation value with respect to $|\varphi_0\rangle$. The explicit form of the above constraints is the following:
\begin{eqnarray} 
  &&{g_{i\sigma}}^2\langle \hat{e}_i\rangle_0 + \sum_{\sigma}{g_{i\sigma}}^2 \langle s_{i\sigma} \rangle_0 + {g_{d,i}}^2\langle \hat{d}_i \rangle_0 = 1,\label{conste}\\
  &&{g_{i\sigma}}^2 \langle s_{i\sigma} \rangle_0  + {g_{d,i}}^2\langle \hat{d}_i \rangle_0 = \nupzr i \label{consts}.
\end{eqnarray}
In the limit of infinite dimensions the effect of the projectors $P_i$ requires the renormalization of the hopping amplitudes between different sites
\cite{MV1,geb}. These renormalization factors can be written as:
\begin{eqnarray}\label{renorm}
 \sqrt{\qup i } =  \frac{g_{e,i}g_{\sigma,i}\sqrt{\langle\hat{e}_i\rangle_0\langle\hat{s}_{\sigma,i}\rangle_0} + g_{d,i}g_{\bar{\sigma},i}\sqrt{\langle\hat{d}_i\rangle_0 \langle\hat{s}_{\bar{\sigma},i}\rangle_0}}
 { \langle n_{i\sigma}\rangle_0(1-\langle n_{i\sigma}\rangle_0) }.\\ \nonumber
\end{eqnarray}
By substituting Eqs.~(\ref{conste}) and (\ref{consts}) into Eq.~(\ref{renorm}) one arrives at an expression for $\sqrt{q}_{i\sigma}$ that is only a function of $g_{d,i}$, $\nupzr i$ and $\ndwzr i$ as:
\begin{equation}
 \sqrt{\qup i } =  \frac{ \sqrt{ (1 - \langle\hat{n}_i\rangle_0 + d_i)(\nupzr i - d_i) } + \sqrt{d_i(\ndwzr i - d_i)}}
 { \langle n_{i\sigma}\rangle_0(1-\langle n_{i\sigma}\rangle_0) },\\ \nonumber
\end{equation}
where $d_i ={g^2_{d,i}} \langle \hat{n}_{i\sigma}\rangle_0\langle\hat{n}_{i\bar{\sigma}}\rangle_0 $ is the probability of double occupancy that is calculated within $|GWF\rangle $ and $\langle\hat{n}_i\rangle_0 = \nupzr i + \ndwzr i$. Moreover, in addition to Eq~.(\ref{renorm}) the relation 
$\langle \nup i \rangle_{gutzwiller} = \langle \nup i \rangle_0$ holds in the limit of infinite dimensions. By considering the above relations the total energy functional of an inhomogeneous system has the following form,
\begin{eqnarray}\label{hexpect0}
\langle \hat{H} \rangle_{GWF} &=& \sum_{\langle ij \rangle, \sigma } -t_{ij}\sqrt{\qup i } \sqrt{\qup j } \langle \cdgiup{i}\ciup{j}  \rangle_0 +\sum_{i,\sigma} v_i \langle \hat{n}_{i\sigma} \rangle_0,  \nonumber\\
 &+& \sum_{i} Ug^2_{d,i} \langle \hat{n}_{i\sigma}\rangle_0\langle\hat{n}_{i\bar{\sigma}}\rangle_0, 
\end{eqnarray}
 The conditions $\langle GWF|GWF\rangle = \langle\varphi_0|\varphi_0\rangle$ and $\langle\varphi_0|\varphi_0\rangle = 1$ are used in the above relation, the first relation itself is a consequence of the infinite dimensional limit
 and the second relation is just the normalization condition for $|\phi_0\rangle$. 

Away from half filling the problem of minimizing the energy functional is combersome because the renormalization factors, $q_{i\sigma}$, are functions of $\nupzr{i}$.  Therefore it is impossible to simply vary the above energy 
functional with respect to $\langle\phi_0|$. A possible approach, similar to DFT, is to start with an arbitrary value for $\nupzr{i}$ and then to expand the energy functional as function of $\nupzr{i}$ up  to linear order around
the starting $\nupzr{i}$. This allows us to vary the energy functional with respect to $\langle\phi_0|$, moreover this variation together with the normalization condition for $|\phi_0\rangle$ leads one to solve an eigenvalue 
problem, and a new value of $\ndwzr{i}$ can be calculated by using the resulted wave function. This should be done until the desired convergence of the wave-function or energy functional is achieved.

However, to avelliate this complication, instead of calculating the expectation value $\langle\hat{n}_{i\sigma}\rangle_0$, we introduce a set of new variational parameters $n_{i\sigma}$s that will play the role of local noninteracting occupancies (local noninteracting
density matrices) which appear in the renormalization factors and double occupancies. It is then possible to let $n_{i\sigma}$ vary independently from $|\phi_0\rangle$. The energy functional that should be optimized 
 has now the following form for a simple cubic slab geometry with periodic boundary conditions in the $x-y$ plane with free $(001)$ surfaces:
\begin{eqnarray}\label{hexpect}
\langle\hat{H}\rangle &&= \sum_{i,k_\|,\sigma}(q_{i\sigma}\epsilon_{k_\parallel}+v_{i})\langle\phi_0|\cdgiup{ik_\|}\ciup{ik_\|}|\phi_0\rangle\nonumber\\
 && \nonumber-\sum_{\langle ij\rangle k_\|\sigma} \sqrt{\qup i}\sqrt{\qup j}t \langle\phi_0|\cdgiup{ik_\|}\ciup{jk_\|} |\varphi_0\rangle\nonumber\\
 && +\sum_{i \sigma}\lambda_{i\sigma}(\sum_{k_\parallel}\langle\phi_0|\cdgiup{ik_\|}\ciup{ik_\|}|\phi_0\rangle -N_{k_\|}{n}_{i\sigma})\nonumber\\
 && +\varLambda(N_{k_\|}\sum_{i\sigma} {n}_{i\sigma}- N ) + E(1-\langle\varphi_0|\varphi_0\rangle)\nonumber\\
 && + \sum_{i} N_{k_\|} U g^2_{d,i}n_{i\sigma}n_{i\bar{\sigma}},
\end{eqnarray}
where $\epsilon_{k_\parallel} = -2t( \cos{k_x}+\cos{k_y} )$, the Lagrange multipliers $\lambda_{i\sigma}$ are introduced to fix $n_{i\sigma}$ to $\langle\hat{n}_{i\sigma}\rangle_0$. $\Lambda$ is introduced to fix the total
number of electrons, $E$ is considered to make sure that $|\varphi_0\rangle$ is normalized, $i$ and $j$ are index of layers in the $z$ direction and $N_{k_\|}=N_{k_x}N_{k_y}$ is the total number of k-points. 
The optimization of the Lagrange function is performed through an iterative procedure, starting with a minimization with respect to $|\varphi_0 \rangle$, which leads to a Schr\"odinger-like eigenvalue problem that has to be solved for each k-point:
\begin{equation}\label{eigen}
\begin{split}
&\sum_{i\sigma}(q_{i\sigma}\epsilon_{k_\parallel} + v_{i} + \lambda_{i\sigma})\cdgiup{ik_\|}\ciup{ik_\|}|\varphi_0\rangle -\\
&\sum_{\langle ij \rangle \sigma} \sqrt{\qup i}\sqrt{\qup j}t\cdgiup{ik_\|}\ciup{jk_\|}|\varphi_0\rangle  = E_{k_\|}|\varphi_0\rangle,\\
\end{split}
\end{equation}

The resulting non-interacting many-body ground state energy and wave-function are computed in the following way: $E_{NI}=\sum_{E_{k_\|,n}<E_F} E_{k_\|,n}$ and $|\varphi_0\rangle = \prod_{E_{k_\|,n}<E_F} \cdgiup{k_\|,n}\cdgidw{k_\|,n}|0\rangle$, where $E_F$ is the Fermi energy and $n$ is the quantum number for the energy level of each k-point. The above non-interacting state $|\varphi_0\rangle$, which is now implicitly a function of  all the variational parameters 
 $\lambda_{i\sigma}$, $n_{i\sigma}$, $g_i$ and $\varLambda$, should be inserted into Eq.~(\ref{hexpect}) which becomes,
\begin{figure}
\begin{center}
 \includegraphics[width=\columnwidth]{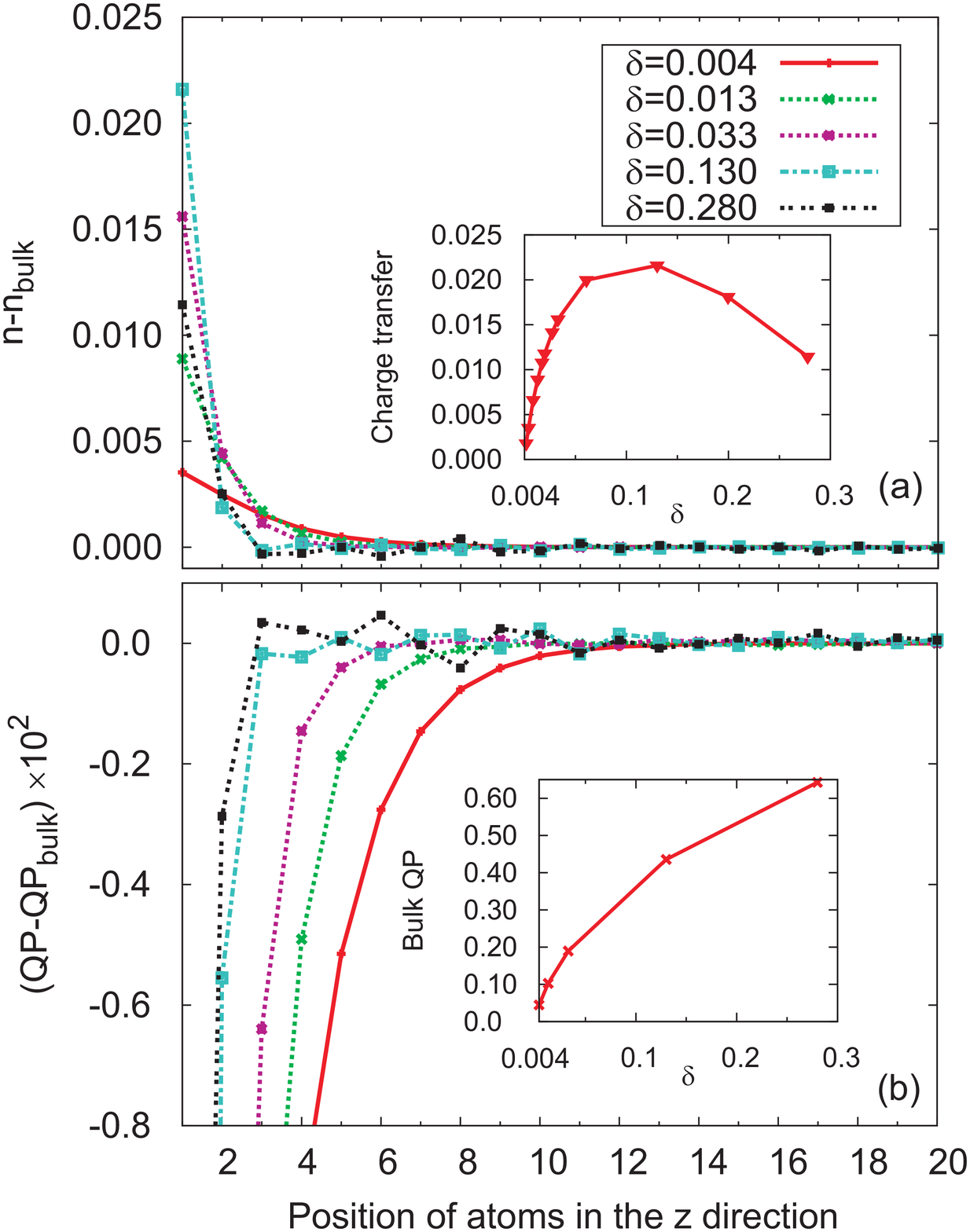}
   \caption {(Color online) (a) Charge distribution for different dopings. Inset: charge transfer from bulk to surface as a function of doping; (b) QP weight relative to the bulk QP weight near the surface for different dopings. 
   Inset shows the doping dependence of the bulk QP weight.}
   \label{fig1}
   \vspace{-1.0cm}
\end{center}
\end{figure}
\begin{eqnarray}
\langle\hat{H}\rangle&=& E_{NI}(n_{i\sigma},\lambda_{i\sigma},g_{d,i} ,|\varphi_0\rangle) - N_{k_\|}\sum_{i,\sigma}\lambda_{i\sigma} n_{i\sigma}\\
&+& \varLambda(N_{k_\|} \sum_{i,\sigma} n_{i\sigma} - N ) + N_{k_\|}\sum_i U g^2_{d,i}n_{i\sigma}n_{i\bar{\sigma}}.\nonumber
\end{eqnarray}
 In the next step we search for the stationary points of the above Lagrange function of a slab geometry for a paramagnetic system with $n_{i\sigma}=n_{i\bar{\sigma}}=n_{i}$ and $\nupzr{i}=\ndwzr{i}$ as:
\begin{widetext}
\begin{eqnarray}
\frac{\partial\langle\hat{H}\rangle}{\partial g_{d,i} } &=& 2\frac{\partial \qup i}{\partial g_{d,i} }(\tilde{t}_i +\delta_{i,j\pm1}\sqrt{\frac{\qup j}{\qup i}} \tilde{t}_{ij}) + 2N_{k_\|} U {n_i}^2g_{i,d}=0,\label{nlst1}\\
\frac{\partial \langle\hat{H}\rangle}{\partial n_{i}} &=& 2\frac{\partial \qup i}{\partial n_{i}}( \tilde{t}_i + \delta_{i,j\pm1}\sqrt{\frac{\qup j}{\qup i}}\tilde{t}_{ij}) + 2N_{k_\|}(\varLambda-\lambda_i) + 2N_{k_\|}  U{g_{i,d}}^2n_i=0,\label{nlst2}\\
\frac{\partial \langle\hat{H}\rangle}{\partial \lambda_{i}} &=& 2\langle\varphi_0|\sum_{k_\|} \cdgiup{ik_\|}\ciup{ik_\|} |\varphi_0\rangle - 2N_{k_\|} n_{i} = 0, \label{nlst3}\\
\frac{\partial \langle\hat{H}\rangle}{\partial \varLambda} &=& ( N_{total} -2N_{k_\|}\sum_{i} n_i ) = 0 \label{nlst4} ,
\end{eqnarray}
\end{widetext}
  where the spin index of renormalization factors and $\lambda_{i\sigma}$ is droped due to paramagnetic condition, $\tilde{t}_i=\sum_{k_\|} \epsilon_{k_\parallel} \langle\varphi_0|\cdgiup{ik_\|}\ciup{ik_\|}|\varphi_0\rangle$, $\tilde{t}_{ij} = -t\sum_{k_\|}\langle\varphi_0|\cdgiup{ik_\|}\ciup{jk_\|}|\varphi_0\rangle$
 and $N_{total}$ is the total number of electrons. $\tilde{t}_{i,i+1}$ and $\tilde{t}_{0,1}$ are equal to zero at edge of the slab. For a detailed derivation of the  saddle point equations for an slab geometry in 
 the paramagnetic case  we refer Ref.~[\onlinecite{nasr}]. This set of nonlinear equations can be solved by using a nonlinear solver based on Newton and/or Quasi-Newton methods. 
 Notice that $|\phi_0\rangle$ is still implicitly a  function of the variational parameters and has to be updated again through Eq.(~\ref{eigen}) during the evaluations of the saddle point equations throughout the optimization procedure. This
 means that we are all the time working with a $|\phi_0\rangle$ which satisfies the condition $\frac{\delta \langle \hat{H} \rangle}{\delta \langle\phi_0|} = 0$.\\ 
It should be noticed that together with the saddle point equations the electrostatic forces due to long-range electron-electron and electron-ion interactions should be in principle considered. However since the back ground 
 permitivity of strongly correlated materials is usually very high\cite{colossal},  we tested the solutions with various high values of background permitivity and observed that long-range screening is negligible\cite{nasr}. 
We therefore set the value of the back ground permitivity to infinity in our calculations and ignore these effects. In order to numerically solve the set of saddle point equations, we use a $16 \times 16$ 
Monkhorst-Pack\cite{monkhorst} k-grid for which the total energy is well converged. We report results for $q_i$ as being the position dependent QP weight, which is a measure of the mobility of the electrons within Fermi liquid
theory. The inverse of the QP  weight is proportional to the mass renormalization which becomes divergent for $q_i=0$ corresponding to an insulating phase\cite{fazekas}. The parameters $U$ and $v$ are scaled by the tight binding
parameter $t$. Throughout this work the thickness of the slab is taken $L_z=90$ in units of the lattice constant.
\section{Results}
\subsection{ Hole doped correlated slab}
\begin{figure}
\begin{center}
   \includegraphics[width=\columnwidth]{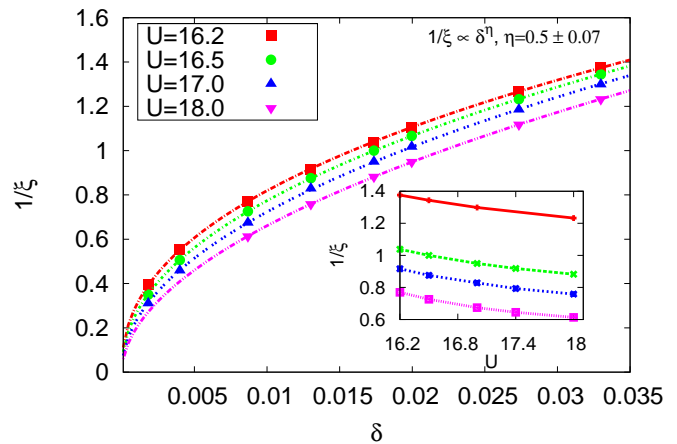}
   \caption {(Color online) (a) Inverse correlation length as function of doping for values of $U=16.2, 16.5, 17.0, 18.0$. The inset shows the inverse of the correlation length as function of $U$ for four values of $\delta$=0.009, 0.013, 0.017, 0.032 from 
   buttom to top curves.}
   \label{fig2}
  \vspace{-1.0cm}
\end{center}
\end{figure}
In Fig.~\ref{fig1}(a) we depict the charge distribution near the surface for different values of doping and $U=16.2$, which is larger than the bulk critical U for the half-filled case, i.e. $U^{bulk,hf}_C=16$. The surface region in 
which the charge density recovers its bulk value is doping dependent, resulting in the doping dependent correlation length. Higher doping corresponds to lower correlation length. In the inset of Fig.\ref{fig1}(a) we present the charge
transfer from the bulk to the surface ($n_{surface}-n_{bulk}$). The doping dependence of this charge transfer is non-monotonic and is maximum around $\delta=0.15$. While our results for the charge transfer are in agreement with recent  DMFT calculations for a hole doped semi-infinite single band Hubbard model\cite{nourafkan} in the limit of large enough doping, our scaling analysis shows that considering only few layers for the QP calculation may not be enough, specially for values of doping near half filling for which the correlation length is larger that 6 lattice constants.
In Fig. ~\ref{fig1}(b) the spatial distribution QP weights $(q_i - q_{bulk})$ are plotted for different values of doping and $U = 16.2$. Like in the half-filled case \cite{potthof,michele,nasr} the QP of electrons near the surface 
sites is suppressed due to the reduced coordination number together with the charge transfer to the surface sites from the bulk, which in turn results in a lack of kinetic energy and an enhancement of correlation effects. One 
can also observe Friedel oscillations which are more pronounced for higher doping due to lower correlation lengths. The inset of Fig.~\ref{fig1}(b) shows the doping dependence of the bulk QP weight, $q_{bulk}$,  which is in 
agreement with previous works and shows that by increasing the doping, correlation effects are weaker \cite{kotliar}. The correlation length can also be extracted from the spatial distribution of the QP weight near the surface. Similar to the 
dependence of the charge density, the QP weight recovers its bulk value within a characteristic length scale that depends on the correlation length. Friedel oscillations can also be observed but are suppressed for lower doping.
\begin{figure}[ttt]
\begin{center}
   \includegraphics[width=\columnwidth]{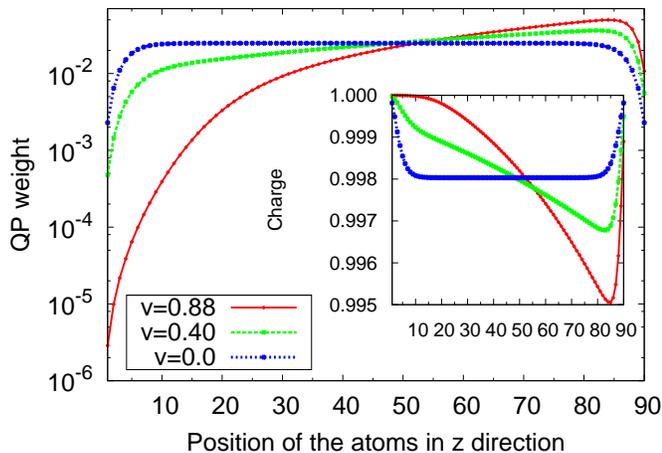}
   \caption {(Color online) QP weight distribution and charge distribution (inset) for $U=16.2$, $\delta=0.002$ and three different values of the electric field.}
   \label{fig3}
   \vspace{-0.5cm}
\end{center}
\end{figure}
Following Ref.~\onlinecite{michele} we observe that the spatial distribution of $\sqrt{q(x)}-\sqrt{q_{bulk}}$ is well fitted by an exponential decay for different values of the Hubbard repulsion and doping:
\begin{equation}
\sqrt{q(x)} = \sqrt{q_{bulk}} + (\sqrt{q_{surface}}-\sqrt{q_{bulk}}) e^{-\frac{1}{\xi}(x-1)},
\end{equation}
where $\xi$ is the correlation length and $x$ the number of layer, starting from $x=1$. Since the correlation length, $\xi$, depends on both $U$ and $\delta$, by fitting separately the spatial distribution of the QP weight we 
extract the corresponding correlation lengths. The results are summarized in Fig.~\ref{fig2}, where $1/\xi$ is plotted as a function of doping for different values of the Hubbard repulsion. We can extract a simple power-law 
dependence for the inverse correlation length: $\frac{1}{\xi} = A{\delta}^\eta$, with a mean-field-like exponent \cite{micheleL,costi}, $\eta = 0.5\pm 0.07$, and a prefactor $A$ that is only a function of $U$. 
The inset of Fig.~\ref{fig2} shows the inverse correlation length as a function of $U$ for different dopings and, as expected, it is enhanced for higher Hubbard repulsions. The power law dependence of the 
correlation length versus doping shows that for half-filling the correlation length diverges which is a signature of the MIT that occurs for $U>U^{bulk,hf}_c$. A similar dependence of the correlation length versus Hubbard
repulsion is observed for half-filling but when $U<U^{bulk,hf}_c$ \cite{micheleL}. In the latter case the criticality is governed by the Hubbard repulsion rather than the doping level.
\begin{figure}
\begin{center}
 \includegraphics[width=\columnwidth]{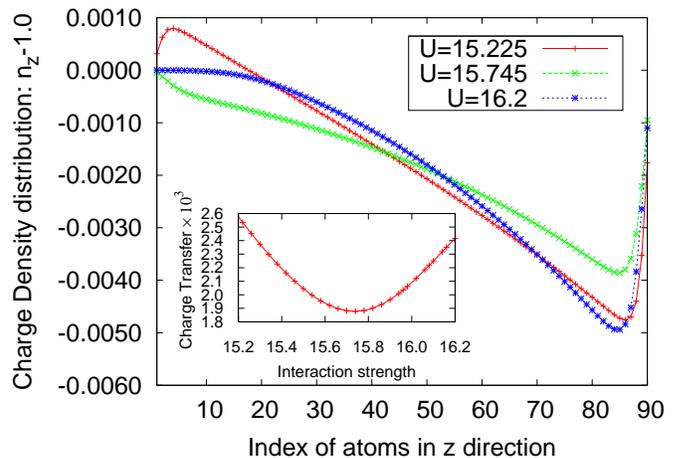}
 \caption {(Color online) Charge distribution; the inset shows the charge transfer as function of $U$ for fixed $v=0.88$ and $\delta=0.002$}
\label{fig4}
\vspace{-1.0cm}
\end{center}
\end{figure}
\subsection{ The effect of electric field}
The effect of an external electric field perpendicular to the slab on the spatial distribution of the QP weight is shown in Fig.~\ref{fig3} for $U=16.2$, $\delta=0.002$ and different values of the voltage difference. 
The inset shows the charge distribution for the same parameters. The main effect of the electric field is to redistribute the charges within the slab, however in the strongly correlated regime when the Hubbard repulsion
exceeds a certain crossover value, correlation effects enhance the charge transfer from less correlated sites to more correlated ones. This correlation enhanced charge redistribution results in the accumulation of charges
near the surface layers, bringing one side of the slab very close to half-filling. This effect is largest for $U>U^{bulk,hf}_c$. 

\begin{figure}
\begin{center}
   \includegraphics[width=\columnwidth]{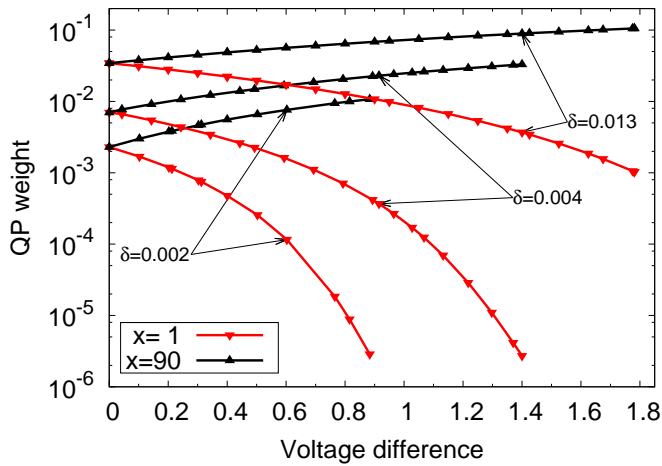}
   \caption {(Color online) QP for x=1 and x=90 as function of voltage difference for $U=16.2$ and three values of $\delta$.}
   \label{fig5}
   \vspace{-1.0cm}
\end{center}
\end{figure}

To better clarify the correlation effects on the surface states of a correlated slab in the 
presence of an electric field we depict in Fig.~\ref{fig4} the charge and quasiparticle(QP) distribution of a slab consisting of $90$ layers thick and a voltage difference $v=0.88$. The charge distribution for
$U=15.22$ shows peaks near the surfaces, as expected, however this behavior disappears for $U=15.74$ and  $U=16.2$. This shows a clear crossover regime related to the enhancement of correlation effects. On the other hand
the naive expectation that the effect of an increased Hubbard repulsion is only to screen out the electric field, fails to explain the behaviour of the system in the presence of the electric field in the strong coupling regime.
As shown in Fig.~\ref{fig4} by increasing the Hubbard repulsion, the charges do not go away from the surface but instead are accumulated at the surface. This mechanism of charge transfer from the places with enhanced 
delocalization to the places with enhanced correlations leads to a non-trivial enhancement of QP difference between the surfaces for large Hubbard repulsions. To further understand the charge redistribution enhancement due to
correlation effects, we present in the inset of Fig.~\ref{fig4} the charge difference between the layers with highest charge density and the layers with lowest charge density as function of $U$. This can be considered as a 
measure of the charge transfer throughout the system. As is clear from the inset of Fig.~\ref{fig4} there is a crossover value for $U$, given a fixed doping $\delta=0.002$ and voltage difference $v=0.88$. Above this value 
the effect of the $U$ plays a different role in the charge redistribution in the system. While below the crossover interaction the Hubbard repulsion competes with $v$ to prevent charge redistribution due to voltage difference,
above the crossover it enhances the charge redistribution in favor of $v$. As is obvious from Fig.~\ref{fig3} the maximum QP weight is already achieved after a few layers from the surface on that side of the slab where the
deviation of the charge density from half-filling is maximal. The reason that the QP weight is not maximal exactly at the surface is because of the suppression of the kinetic energy near the surface. On the other side of the 
slab, for larger electric fields the charge transfer assures that the charge density is near half-filling. Therefore, due to local correlation effects the QP weight is strongly suppressed. While the charge density near the 
surface is very close to half-filling (i. e. $n-1\simeq 10^{-7}$) one may infer that the residual QP indicated in Fig. \ref{fig3} for $x=1$ is mostly due to the proximity of the surface site to sites with higher QP weight rather
than due to the local doping effect of these regions. In order to better understand the dependence of the QP weight on opposite sides of the slab, in Fig.~\ref{fig5} we show the QP weight for layers $x=1$ and $x=90$ as function
of voltage difference for three different values of doping.\\
The QP weights on the two surfaces differ by \textit{many orders of magnitude}. For larger doping, higher electric fields are needed in order to achieve the same QP weight difference. This is because of the competing influence of
doping and Hubbard repulsion on the correlation effects. The huge difference in QP weight near the two surfaces could be used for creating a transistor-like device made of strongly correlated materials. By using the surface states
to conduct current one can simply  switch on/off the device by switching the polarity of the gate. Thus, turning on/off the electric conduction is now a consequence of the surface resistance ratio of the two sides.

\section{Conclusions} 
By using an inhomogeneous Gutzwiller approach applied to the paramagnetic single band Hubbard model for a slab geometry we described a hole doped Mott thin-film. In the absence of applied electric field we calculated 
the position dependent charge density and QP weight and showed that the inverse correlation length has a power law dependence on doping.\\
When a perpendicular electric field is applied, charges will accumulate on one side of the slab. This correlation enhanced charge redistribution will in turn induce a large difference in the QP weight on the two sides
of the slab, which was found to be as large as five orders of magnitude. We propose that a three terminal device with surface contacts can take advantage of this effect. For resistance switching purposes one would expect
large on/off ratios of surface resistances when the electric field switches polarity.
\begin{acknowledgements}
 This work was supported by the Flemish Science Foundation (FWO-Vl). One of us (LC) is a postdoctoral fellow of the FWO-Vl.
\end{acknowledgements}

\end{document}